# The Proposal of a Novel Software Testing Framework

Munib Ahmad [1], Fuad Bajaber [2], M. Rizwan Jameel Qureshi [2]

[1.] Department of Computer Science, COMSATS Institute of Information Technology, Lahore, Pakistan
[2.] Faculty of Computing and Information Technology, King Abdulaziz University, Jeddah, Saudi Arabia
munib_kamboh@ciitlahore.edu.pk, fbajaber@kau.edu.sa, rmuhammd@kau.edu.sa

**Abstract:** Software testing is normally used to check the validity of a program. Test oracle performs an important role in software testing. The focus in this research is to perform class level test by introducing a testing framework. A technique is developed to generate test oracle for specification-based software testing using Vienna Development Method (VDM++) formal language. A three stage translation process, of VDM++ specifications of container classes to C++ test oracle classes, is described in this paper. It is also presented that how derived test oracle is integrated into a proposed functional testing framework. This technique caters object oriented features such as inheritance and aggregation, but concurrency is not considered in this work. Translation issues, limitations and evaluation of the technique are also discussed. The proposed approach is illustrated with the help of popular triangle problem case study.

## 1. Introduction

Software testing is really a difficult job. Recently software testers have increased their dependency on automated testing to test software. Software test automation is often difficult and complex process. Generating and running the test cases and generating and verifying test results are very important in test automation. We identify the needs to be verified during designing of a test. A set of expected results are required to verify the actual results for each test. The process of expected results generation is done using test oracle (Binder, 2002). Expected result generator and comparator are two main components of a test oracle. Implementation under test is used to generate actual result for a particular test case. Then actual result is compared with expected result, generated by expected result generator for evaluation whether actual result is correct. The output of comparator will be 0 or 1. If actual and expected results are same then ok, otherwise error will be declared.

People are rapidly adopting and relying on software to perform their daily activities. This level of dependency and confidence in software requires the checking of its correct behavior for safety of the people and their business. Correctness of the behavior of software depends on how much level you are performing the testing of that software (Peters and Parnas, 1998).

Systematic testing especially supported by suitable tools can greatly increase the effectiveness of system verification and the confidence in the correct functioning of the system (Takahashi, 2001). Automated testing has the ability to reduce the testing time and save up to 80% of testing costs because automated tests can execute test cases much faster than manual testing (Takahashi, 2001). All software testing researches and practices assume that there is some mechanism, an oracle, for determining whether or not the output from a program is correct. "A Perfect Oracle would be behaviorally equivalent to the implementation under test (IUT) and completely trusted. In effect, it would be a defect free version of the IUT. It would accept every input specified for the IUT and would always produce a correct result." (Binder, 2002). Therefore, the development of a perfect oracle will be as difficult as the development of the original software.

The rest of the paper is organized as: section 2 discussed related work. Problem definition is given in section 3. The proposed technique and its limitations are discussed in section 4. In section 5, a case study of popular triangle problem is conducted to validate our proposed research. The evaluation of proposed technique is described in section 6.

## 2. Related Work

The test oracle generation for specification-based software testing techniques can be classified on the basis of formal specification notations. We can categorize formal specifications into six categories i.e., model-based, algebraic, Logic-based, Net-based/Graphical, Process Algebra, and tabular/equation execution-based. An emphasis is given on the research contributions that target test oracle generation for specification-based software testing.

The main idea behind this classification is to ensure that most of the major contributions and main techniques should be covered and presented in a manner that can lead to a comparison based study of the research efforts that are being carried out in the area of test oracle generation for specification-based software testing. So this classification is not mean to be exhaustive. Table 1 summarizes our classification of test oracle generation for specification-based software testing techniques.

Table 1. Classification of Techniques

| Specification Category | Specification | Techniques and Contributions |
|---|---|---|
| Model-based | Z | Stocks and Carrington (1996), Horcher (1995) |
| | Object Z | Carrington et al. (2000) |
| | VDM-SL | Meudec (1998) |
| | JML (Java Modeling Language) | Boyapati et al. (2002) |
| | State-based | Blackburn and Busser (1996) |
| Algebraic | LOBAS | Doong and Frankl (1991) |
| Logic-based | HLTL, GIL, ITL | Dillon and Ramakrishna (1996) |
| Tabular/Equation execution-based | Anna | Hagar and Bieman (1996) |

Test oracle generation techniques for specification-based software testing can be evaluated on the basis of seven points i.e. notation independence, object-orientation, coverage, accuracy of information, usability, complexity, temporal relationship, automation and tool support on the basis of their importance and criticality in the development of an automated test oracle generation for specification-based software testing.

Few researchers (Richardson, 1994) and (Dillon and Ramakrishna, 1996) presented notation independence techniques which are not strictly dependent on the syntax and structure of a particular formal specification notation. Carrington et al. (2000) and Boyapati et al. (2002) targeted the object oriented features. A technique reveals most number and types of faults as it may have sufficient coverage of formal specification. Stocks and Carrington (1996), Doong and Frankl (1991), Hagar and Bieman (1996), and Boyapati et al. (2002) presented test oracle generation techniques to provide sufficient coverage of formal specifications. Most of the researchers generated expected result generator and comparator in their research. Horcher (1995) developed a technique in which Z specifications may be used instead of a test oracle to validate the observed test results automatically.

Test oracle provides accurate information that becomes more important for the testing of safety critical software. This is because we cannot afford faults in such kind of systems. Test oracle techniques provide accurate information, are presented in (Horcher, 1995). Test oracle and system under test (SUT) can be used in parallel to test the intended behavior of SUT and test oracle should provide results in useful manner, for examples in the form of bits and bytes (True or False, 0 or 1), electronic signals, hardcopy and display (Binder, 2002). This will improve usability of test oracle. Test oracles generated by Horcher (1995) provide information in useful manner and can be used in parallel with SUT. Blackburn and Busser (1996), and Horcher (1995) developed techniques for safety critical systems and these techniques are very complex in nature. Test oracle generated for testing of real time systems should generate the results in specified time.

### 3. Problem Definition

VDM++ is a popular formal specification language in software industry during last several years. It supports object oriented features and provides full specifications coverage. Meudec (1998) discussed a technique to generate test cases from specifications written in VDM-SL. Meudec (1998) did not address expected result generation in his work; in other words Meudec did not address test oracle generation using formal specifications written in VDM-SL. This work has no support for object oriented features. There is a need to develop test oracle using VDM++ formal specifications to support object oriented features and having full coverage of formal specifications. Test oracle can also provide accuracy of information and easily be used in parallel with SUT.

### 4. The Proposed Testing Framework

In this research we are focusing on the methodical derivation of active test oracles from formal object-oriented specifications. Using a number of specifications of container classes, we have produced a mapping from VDM++ specifications to C++ test oracles. Our aim is that the derived oracles will be general enough to be usable in most testing frameworks. Overall flow graph of our proposed technique and complete testing framework are presented in figure 1.

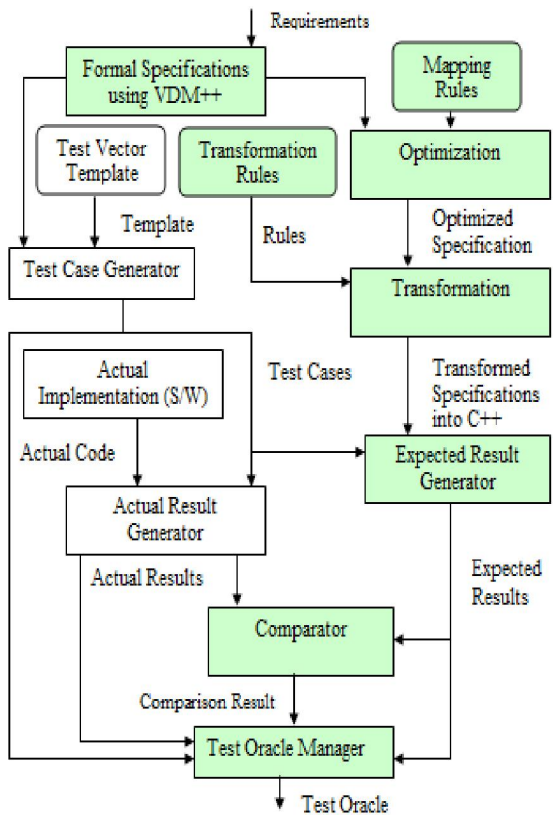

Figure 1. Framework for Specification-based Software Testing

In figure 1, rectangles with circular edges representing document or template and straight rectangles representing phases of given framework in which some processing is being done. The colored part of this framework is relating to our proposed technique. In this part, formal specifications written in VDM++ are provided to generate test oracle for specification-based software testing. The first phase of our proposed technique is to optimize these specifications using mapping rules in such a way that these specifications can easily be transformed into C++ test oracle classes. Optimization transforms the specification to a form more suited to systematic translation to C++. In next phase i.e. transformation phase structural mapping is performed to produce C++ code with structure corresponding to that of the specification. Then for mapping of predicates to C++ code translation is performed in this phase. Expected Result Generator is used to produce expected results using this C++ code produced from VDM++ specifications. Test cases produced manually or systematically using VDM++ specifications applying any test case generation technique are provided to Expected Result Generator to produce expected results. Test case generation is not a part of our proposed technique.

Actual Result Generator produces actual results using Implementation under Test (IUT). Same test cases are provided to Actual Result Generator and Expected Result Generator to produce actual and expected results respectively. Actual Result Generator is not a part of our proposed technique. Actual Results can be produced manually, but in systematic testing Actual Result Generator can be automated to automate the whole testing process. These actual and expected results are provided to comparator to compare. At the end Test Oracle Manager manages the test cases, actual and expected results and their comparison results. This information will be useful in regression testing as well as for documentation. Detail discussion of all the phases included in our proposed technique i.e. Optimization, Transformation, Expected Result Generator, Comparator, and Test Oracle Manager is presented in following subsections.

**4.1 Optimization**

Optimization is the rearrangement of the specification to simplify translation to an implementation language. We performed optimization in two steps. In first optimization step, we mapped VDM++ data types for declared variables to its equivalent C++ data types according to the problem for which VDM++ specification is written. In second optimization step, we convert VDM++ classes into its corresponding C++ classes.

**4.1.1 Step1: Mapping of Data Types**

VDM++ data types can be divided into two categories i.e. Basic and Compound data types. Mapping of Basic data types into its equivalent C++ data types is given in Table 2.

Table 2. Mapping of Basic Data Types

| Data type in VDM++ | Equivalent data type in C++ |
|---|---|
| Boolean | Boolean |
| nat1, nat, int | Int |
| rat, real | Float |
| Char | Char |
| Quote | Enum |
| Token | Vector of type string |

There are eight compound types in VDM++ i.e. Set, Sequence, Map, Product, Composite, Union and Optional, and Function types (Meudec, 1998). Mapping of compound data types can be performed as:

- Set data type can be mapped in C++ as compound data type set using Standard Template Library (STL). We implemented functions in C++ in STLSetAlgos.h

- Standard template library for those operators of Set data type in VDM++ whose corresponding functions are not provided by C++ e.g. Union, intersection, difference, subset, proper subset.
- To map Sequence data type in C++, declare list of the values of the same type as the sequence elements. Then these elements assigned to a vector type variable with the same name as it is used in VDM++ specification. We implemented functions for tail, elements, indexes, and concatenation operators in C++ in STLSequenceAlgos.h Standard template library. We extended STL vector and Set libraries in C++ to implement these operations.
- Map type can be mapped in C++ using map type and we extended STL map and Set libraries by implementing STLMapAlgos.h Standard Template Library in C++ to implement the operations provided by VDM++ for Map data type.
- Product and Composite types in VDM++ can easily be mapped in C++ using struct type. All operations of Product and Composite types are same.
- Union and optional type is a bad practice (Meudec, 1998), so tester can decompose this type into those relevant C++ data types for which this union and optional type contains the elements after understanding the specification. So no particular mapping rule can be provided for union and optional type. Its mapping in C++ is totally depending on tester's experience and his/her specification understanding.
- Function type in VDM++ can be mapped in C++ by implementing expression for the body of the function type. Then result of this expression is assigned to the variable of the same name as the name of the variable of this function type in specification and the data type of this variable should be same as the resultant type of this expression.

### 4.1.2 Step2: Mapping of Classes

Semantics of class header are same in C++ as VDM++, but syntax in C++ is different. Structural mapping rules of VDM++ class body (optional) in C++ can be performed as:

- A set of value definitions (constants) can be transformed in C++ with const declaration using same name and access specifier i.e., public or protected as specified in the specification.
- A set of type definitions can be transformed as discussed in section 4.1.1, but type definition can be public or protected as specified in the specification.
- Function definitions: Semantics of function definitions are same in C++ as VDM++. While transforming in C++ public, private, and protected functions specified in VDM++ are mapped as public, private, and protected respectively. Explicit, implicit, and extended explicit definitions of functions in VDM++ specification can be transformed manually in C++ after understanding the specification. The difference between them is just the level of abstraction of the input and output parameter definitions. Pre-condition expression can be implemented and ensured in C++ before calling of this function. Post condition expression can be implemented and ensured in C++ at the end of the specified function body. Polymorphic functions in VDM++ specification can be transformed in C++ using function overloading. Higher order functions in specification can be implemented in C++ using function recursion.
- A set of instance variable definitions can be transformed in C++ by declaring these variables public or protected as specified in the specification at class level and if these variables are initialized with some value then initialization will be performed in class constructor.
- A set of operation definitions that can act on the internal state will be transformed in C++ by implementing class methods declaring public.
- Transformation rules of the synchronization and thread definitions will be presented in future research.
- Semantics of inheritance are same in C++ as in VDM++, but a little bit difference in syntax is there. This can be implemented using transformation rules, discussed in next section.
- In C++, we implemented a method 'inv()' for the implementation of invariant. This method returns true if invariant is true, and false otherwise. We implemented class 'error' to deal with exception; if invariant becomes false, exception will be thrown. Before and after performing any operation of the class, we have to check this invariant.

## 4.2 Transformation

Transformation is performed by providing transformation rules to transform VDM++ statements into its equivalent C++ statements and providing predicate translation rules. Structural mapping produces skeleton C++ code with structure corresponding to that of the specification, while predicate translation maps predicates to code. The implementation of mapping and transformation rules is illustrated using popular case study of triangle problem, presented in coming section.

## 4.3 Expected Result Generator

To implement Expected Result Generator to produce expected results using C++ code generated from specifications written in VDM++, we implemented 'driver' class. Test cases are provided to Expected Result Generator to produce expected results.

## 4.4 Comparator

To implement comparator, we implement a method with name 'comparator' in 'driver' class to compare the actual and expected results. If actual and expected results match this method returns true and false otherwise.

## 4.5 Test Oracle Manager

Test Oracle Manager manages the test cases, actual and expected results and their comparison results. Tester will provide a template to Test Oracle Manager to manage this information.

## 4.6 Limitations of the Proposed Technique

In this proposed technique concurrency is not addressed. We are working on the automation of our proposed technique to automate the whole process of test oracle generation. We are also working in this direction that how can we minimize the human intervention in this process. It will be the responsibility of the tester to implement driver class to deal with expected result generator.

## 5. Case Study Validation

To illustrate our proposed technique, we use modified version of Mayer's VDM-SL specification of Triangle problem written in VDM++, presented in Table 3 (Meudec, 1998). In this specification a 'Triangle' class is specified, in which three sides of a triangle are taken to judge that the triangle taken by user is equilateral, isosceles, scalene, or an invalid. The measurement of triangle sides is taken using a sequence 'Triangle_sides' of integer type as specified in specification using N*.

Class invariant is specified in which two properties of a triangle are ensured. First property is that the sides of triangle must be three. Second property is that the perimeter of the triangle must be greater than the double of its any side. To find perimeter of the triangle sum of the triangle sides is required. To find sum of the triangle sides a function with name 'sum' is specified which takes a sequence of natural numbers and return a natural value.

To check whether the triangle is equilateral, isosceles, or scalene, a function is specified with name 'variety' which takes a sequence 'Triangle_sides' and returns 'Triangle_type'. 'Triangle_type' is a quote type specified globally in the specification. Now to check whether triangle is valid or invalid another method is specified with name 'classify' which takes a sequence of natural numbers and returns 'Triangle_type'.

Now we generate test oracle using our proposed technique. In the first step of our technique, optimization of data types and classes is performed. In second step, we transformed VDM++ statements and predicate translation is performed. Then to accommodate this test oracle with testing environment a test oracle driver is written.

## 5.1 Optimization

In this step, we rearrange the specification to simplify translation to an implementation language. For 'Triangle_type' which is a quote type will convert to enumerated type in C++. For 'Triangle_sides', which is a sequence of natural numbers will convert to a vector of int type. Optimization for class 'Triangle' is to convert it in C++ with name 'Oracle_Triangle'. Methods of 'Triangle' class in specification are mapped to 'Oracle_Triangle' class in C++ with same name as in specification. Method 'sum' in specification accepts sequence of natural numbers and returns natural number. Now it is mapped in C++ in this way that it will accept vector of int type and return int. Method 'variety' in specification accepts sequence 'Triangle_sides' of natural numbers and returns 'Triangle_type' which is quote type. Now it is mapped in C++ in this way that it will accept vector with name 'Triangle_sides' of int type and return enumerated type 'Triangle_type'. Method 'classify' in specification accepts sequence of natural numbers and returns 'Triangle_type' which is quote type. Now it is mapped in C++ in this way that it will accept vector of int type and return enumerated type 'Triangle_type'. Class invariant is specified with keyword 'inv' in specification is implemented as a method in test oracle with name 'inv'.

Table 3. VDM++ Specification of Triangle Problem

```
triangle_type = INVALID | EQUILATERAL | ISOSCELES | SCALENE
Class Triangle
private triangle_sides = N*
Inv Triangle (sides) = = len sides = 3 ∧ let perim = sum (sides) in ∀ i ∈ elems sides.2*i < perim
functions
        private sum : N* → N
                sum (seq) = = if seq = [ ] then 0 else hd seq + sum (tl seq)
        private variety : Triangle_sides → Triangle_type
                variety (sides) = = cases card (elems sides) of
        1 → EQUILATERAL
        2 → ISOSCELES
        3 → SCALENE
        end
        public classify : N* → Triangle_type
                classify (sides) = = if is_Triangle (sides) then variety (sides) else INVALID
End Triangle
```

The class invariant accepts a sequence of natural numbers in specification and invariant always returns Boolean value. While mapping in C++, 'inv' method of 'Oracle_Triangle' class accepts vector of int type and returns Boolean value. Declarations of 'Oracle_Triangle' class is presented in Table 4.

**5.2 Transformation**
In this step, statements of the VDM++ are transformed into C++ statements and predicate translation is also performed according to the transformation rules. After applying transformation rules, our 'Oracle_Triangle' class is presented in Table 5.

**5.3 Incorporating the Derived Oracle in Testing Framework**
To accommodate our test oracle in the complete testing framework, tester will have to write oracle driver. It is the responsibility of the tester to implement oracle driver class to set an environment to test the actual behavior of component under test and results are compared with the expected results generated by the test oracle. We implemented an oracle driver class with name 'driver' and its declaration is presented in Table 6.

Table 4. Declarations for Oracle_Triangle

```
enum triangle_type {INVALID,
   EQUILATERAL,ISOSCELES, SCALENE};
class Oracle_Triangle {
private: vector<int> triangle_sides;
        triangle_type t;
        bool inv(vector<int> sides);
```

Table 5. Oracle_Triangle implementation

```
        else return hd(seq,1)+sum(tl(seq));}
  bool perim(vector<int> sides) {
            vector<int>::iterator iter;
            iter=sides.begin();
            while(iter!=sides.end()) {
                if(*iter*2<sum(sides)) iter++;
                else return false; }
            return true; }
  triangle_type variety(vector<int> sides) {
   switch((elems(sides,1).size())) {
    case 1:cout<<"\nEQUILATERAL"; return 1;
    case 2: cout<<"\nISOSCELES"; return 2;
    case 3: cout<<"\nSCALENE"; return 3; } }
  public:
   triangle_type classify(vector<int> sides) {
            if (inv(sides)) variety(sides);
 else  {cout<<"\n INVALID"; return 0;
```

Table 6. Oracle Driver Class Declaration

```
class driver {
    public:
        bool comparator(vector<int> sides);
    private:
Oracle_Triangle oracle;
Triangle iut; };
```

Table 7. Driver Class implementation

```
class driver {
  public: bool comparator(vector<int> sides) {
    if(ot.classify(sides)==t.classify(sides))
return true;
return false; }
    private: Oracle_Triangle ot;
    Triangle t; };
```

Table 8. Test cases and their corresponding results generated by our Test Oracle and Meudec (1998)

| ID. | Test Input | Result | Comparator Result | ID. | Test Input | Result | Comparator Result |
|---|---|---|---|---|---|---|---|
| 1 | [0,0,0] | Invalid | True | 19 | [2,3] | Invalid | True |
| 2 | [0,1,1] | Invalid | True | 20 | [4,4,4,4] | Invalid | True |
| 3 | [1,0,1] | Invalid | True | 21 | [M,M,1] | Isosceles | True |
| 4 | [1,1,0] | Invalid | True | 22 | [M,M,M] | Equilateral | True |
| 5 | [3,1,2] | Invalid | True | 23 | [M+1,M-1,M] | Scalene or Invalid | True |
| 6 | [1,3,2] | Invalid | True | 24 | [1,1,1] | Equilateral | True |
| 7 | [2,1,3] | Invalid | True | 25 | [1,2,2] | Isosceles | True |
| 8 | [1,2,5] | Invalid | True | 26 | [2,1,2] | Isosceles | True |
| 9 | [5,2,1] | Invalid | True | 27 | [2,2,1] | Isosceles | True |
| 10 | [2,5,1] | Invalid | True | 28 | [3,2,2] | Isosceles | True |
| 11 | [5,1,1] | Invalid | True | 29 | [2,3,2] | Isosceles | True |
| 12 | [1,5,1] | Invalid | True | 30 | [2,2,3] | Isosceles | True |
| 13 | [1,1,5] | Invalid | True | 31 | [2,3,4] | Scalene | True |
| 14 | [1,2,-6] | Invalid | True | 32 | [3,2,4] | Scalene | True |
| 15 | [-2,-2,-2] | Invalid | True | 33 | [3,4,2] | Scalene | True |
| 16 | [2,2.3,2] | Invalid | True | 34 | [4,3,2] | Scalene | True |
| 17 | ['A',2,3] | Invalid | True | 35 | [4,2,3] | Scalene | True |
| 18 | ['A','A','A'] | Invalid | True | 36 | [2,4,3] | Scalene | True |

At this moment, we suppose that the programmer implemented this class Triangle specified in the specification with the same name. To test Triangle class, test cases required only a sequence of integer type and results are in a Triangle type which is an enumerated type. In driver class, we implemented a method 'comparator' to compare the actual results generated by the implementation under test with expected results generated by the test oracle. 'Comparator' method takes vector of integer type (test case) and returns Boolean value. If actual and expected results match, 'comparator' will return true and false otherwise. If result is false then there is a possibility of error in implementation under test. The complete code of our 'driver' class is presented in Table 7.

**6. Evaluation of the Proposed Technique**

In order to evaluate the efficiency and effectiveness of our proposed technique, we adopted test cases generated by Meudec (1998) from specifications for North's Triangle problem written in VDM-SL and followed structured approach. We modified this VDM-SL specification of North's Triangle problem in VDM++ and followed object-oriented approach.

We used same test cases generated by Meudec (1998) to evaluate the efficiency and effectiveness of our proposed technique because Meudec proof that this test set is adequate for this problem. Remember that test case generation is not a part of our research. Meudec generated thirty six test cases for this problem. This test set is applied to the test oracle produced using our proposed technique to generate expected results. Test cases and their result generated by Meudec for the North's Triangle problem and results produced by our test oracle are presented in Table 8. All the results are same as the results generated by Meudec (1998). Results shows that all the results produced by the test oracle generated using our proposed technique are correct. This evaluation shows that our technique can be applied to generate test oracle for specification-based software testing.

Our proposed technique to generate test oracle from VDM++ specifications is notation dependent and follows completeness. Completeness means our technique supports all three phases of oracle generation process i.e., function generation, expected result generation, and comparator. Our technique supports object orientation while most the techniques presented in the survey do not support object orientation. Our proposed technique provides support to test the intended behavior of class methods as well as interactions between them. Most of the techniques presented in the survey focus on the intended behavior of the functional components in the structural paradigm. Thus our technique provides complete specification coverage.

## 7. Conclusion and Future Work

In this research, we proposed a technique to generate test oracle from VDM++ specifications for specification-based software testing. We used VDM++ formal language because it is widely used in the industry. Test oracle is generated in C++. Most of the test oracle generation techniques in the literature do not support object oriented paradigm and all the phases of test oracle generation process i.e., function generator, expected result generator, and comparator. Our proposed technique supports object oriented paradigm and all the phases of test oracle generation process. Test oracle generated using our proposed technique can also be used in parallel with implementation under test to compare the actual and expected results. This will reduce the testing time and effort.

Future work includes the implementation of concurrency and the complete automation of our proposed technique. Ignorable human intervention will be required after the completion of the automation of this technique. More experimental evaluation is also required to gain high confidence in the software testing using our proposed technique.

**Corresponding Author:**
Dr. M. Rizwan Jameel Qureshi
Department of Information Technology
Faculty of Computing & Information Technology,
King Abdulaziz University
Jeddah 21589, Saudi Arabia
E-mail: rmuhammd@kau.edu.sa

9/26/2013